\input harvmac.tex
\noblackbox
%


\def\unlockat{\catcode`\@=11}
\def\lockat{\catcode`\@=12}

\unlockat

\def\newsec#1{\global\advance\secno by1\message{(\the\secno. #1)}
\global\subsecno=0\global\subsubsecno=0\eqnres@t\noindent
{\bf\the\secno. #1}
\writetoca{{\secsym} {#1}}\par\nobreak\medskip\nobreak}
\global\newcount\subsecno \global\subsecno=0
\def\subsec#1{\global\advance\subsecno
by1\message{(\secsym\the\subsecno. #1)}
\ifnum\lastpenalty>9000\else\bigbreak\fi\global\subsubsecno=0
\noindent{\it\secsym\the\subsecno. #1}
\writetoca{\string\quad {\secsym\the\subsecno.} {#1}}
\par\nobreak\medskip\nobreak}
\global\newcount\subsubsecno \global\subsubsecno=0
\def\subsubsec#1{\global\advance\subsubsecno by1
\message{(\secsym\the\subsecno.\the\subsubsecno. #1)}
\ifnum\lastpenalty>9000\else\bigbreak\fi
\noindent\quad{\secsym\the\subsecno.\the\subsubsecno.}{#1}
\writetoca{\string\qquad{\secsym\the\subsecno.\the\subsubsecno.}{#1}}
\par\nobreak\medskip\nobreak}

\def\subsubseclab#1{\DefWarn#1\xdef
#1{\noexpand\hyperref{}{subsubsection}%
{\secsym\the\subsecno.\the\subsubsecno}%
{\secsym\the\subsecno.\the\subsubsecno}}%
\writedef{#1\leftbracket#1}\wrlabeL{#1=#1}}
\lockat


\def\boxit#1{\vbox{\hrule\hbox{\vrule\kern8pt
\vbox{\hbox{\kern8pt}\hbox{\vbox{#1}}\hbox{\kern8pt}}
\kern8pt\vrule}\hrule}}
\def\mathboxit#1{\vbox{\hrule\hbox{\vrule\kern8pt\vbox{\kern8pt
\hbox{$\displaystyle #1$}\kern8pt}\kern8pt\vrule}\hrule}}
%
\def\exercise#1{\bgroup\narrower\footnotefont
\baselineskip\footskip\bigbreak
\hrule\medskip\nobreak\noindent {\bf Exercise}. {\it #1\/}\par\nobreak}
\def\endexercise{\medskip\nobreak\hrule\bigbreak\egroup}
%

%
%
%

\def\CH{{\cal H}}

\def\CM{{\cal M}}
\def\CN{{\cal N}}
\def\CS{{\cal S}}
\def\CP{{\cal P}}

\def\IZ{\relax\ifmmode\mathchoice
{\hbox{\cmss Z\kern-.4em Z}}{\hbox{\cmss Z\kern-.4em Z}}
{\lower.9pt\hbox{\cmsss Z\kern-.4em Z}}
{\lower1.2pt\hbox{\cmsss Z\kern-.4em Z}}\else{\cmss Z\kern-.4em
Z}\fi}
\def\inbar{\,\vrule height1.5ex width.4pt depth0pt}
\def\IB{\relax{\rm I\kern-.18em B}}
\def\IC{\relax\hbox{$\inbar\kern-.3em{\rm C}$}}
\def\ID{\relax{\rm I\kern-.18em D}}
\def\IE{\relax{\rm I\kern-.18em E}}
\def\IF{\relax{\rm I\kern-.18em F}}
\def\IG{\relax\hbox{$\inbar\kern-.3em{\rm G}$}}
\def\IH{\relax{\rm I\kern-.18em H}}
\def\II{\relax{\rm I\kern-.18em I}}
\def\IK{\relax{\rm I\kern-.18em K}}
\def\IL{\relax{\rm I\kern-.18em L}}
\def\IM{\relax{\rm I\kern-.18em M}}
\def\IN{\relax{\rm I\kern-.18em N}}
\def\IO{\relax\hbox{$\inbar\kern-.3em{\rm O}$}}
\def\IP{\relax{\rm I\kern-.18em P}}
\def\IQ{\relax\hbox{$\inbar\kern-.3em{\rm Q}$}}
\def\IR{\relax{\rm I\kern-.18em R}}
\font\cmss=cmss10 \font\cmsss=cmss10 at 7pt
\def\IZ{\relax\ifmmode\mathchoice
{\hbox{\cmss Z\kern-.4em Z}}{\hbox{\cmss Z\kern-.4em Z}}
{\lower.9pt\hbox{\cmsss Z\kern-.4em Z}}
{\lower1.2pt\hbox{\cmsss Z\kern-.4em Z}}\else{\cmss Z\kern-.4em Z}\fi}
\def\IGa{\relax\hbox{${\rm I}\kern-.18em\Gamma$}}
\def\IPi{\relax\hbox{${\rm I}\kern-.18em\Pi$}}
\def\ITh{\relax\hbox{$\inbar\kern-.3em\Theta$}}
\def\IOm{\relax\hbox{$\inbar\kern-3.00pt\Omega$}}

\def\inbar{\,\vrule height1.5ex width.4pt depth0pt}

\font\cmss=cmss10 \font\cmsss=cmss10 at 7pt
\def\IR{\relax{\rm I\kern-.18em R}}

\lref\aft{I. Antoniadis, S. Ferrara, and T. Taylor, ``${\CN}=2$ heterotic
superstring
and its dual theory in five dimensions," hep-th/9511108, Nucl. Phys. {\bf B
460} (1996) 489.}
\lref\kv{S. Kachru and C. Vafa, ``Exact results for ${\cal N}=2$
compactification of
heterotic strings," hep-th/9505105, Nucl. Phys. {\bf B 450} (1995) 69. }
\lref\serone{M. Serone, ``${\cal N}=2$ Type I-heterotic duality and higher
derivative $F$-terms,"  hep-th/9611017, Phys. Lett. {\bf B 395} (1997) 42.}
\lref\agnt{
I. Antoniadis, E. Gava, K. Narain and T.R. Taylor, ``Topological
amplitudes in string theory,'' hep-th/9307158,
Nucl.Phys. B413 (1994) 162-184; ``${\cal N}=2$
type II-heterotic
duality and higher-derivative $F$-terms," hep-th/9507115,
Nucl. Phys. {\bf B 455} (1995) 109.}
\lref\hmalg{J.A. Harvey and G. Moore, ``Algebras, BPS states, and strings,"
hep-th/9510182,
Nucl. Phys. {\bf B 463} (1996) 315.}
\lref\bor{R.E. Borcherds, ``Automorphic forms with singularities on
Grassmannians," alg-geom/9609022.}
\lref\dkl{L. Dixon, V. Kaplunovsky and J. Louis, ``Moduli dependence of string
loop corrections to gauge coupling constants," Nucl. Phys. {\bf B 355} (1991)
649.}
\lref\gv{D. Ghoshal and C. Vafa, ``$c=1$ string as the topological theory of
the conifold," hep-th/9506122, Nucl. Phys. {\bf B 453} (1995) 121.}
\lref\bcov{M. Bershadsky, S. Cecotti, H.Ooguri and C. Vafa, ``Holomorphic
anomalies in
topological field theory," Nucl. Phys. {\bf B 405} (1993) 279.}
\lref\ks{M. Bershadsky, S. Cecotti, H.Ooguri and C. Vafa, ``Kodaira-Spencer
theory of
gravity and exact results for quantum string amplitudes," hep-th/9309140, Comm.
Math. Phys. {\bf 165} (1994) 311. }
\lref\latst{W. Lerche, A.N. Schellekens and  N.P. Warner, ``Lattices and
strings," Phys. Rep. {\bf 177} (1989) 1.}
\lref\fone{G.L. Cardoso, G. Curio, D. L\"ust and T. Mohaupt, ``Instanton
numbers and
exchange symmetries in ${\CN}=2$ string pairs," hep-th/9603108, Phys. Lett.
{\bf B 382} (1996) 241.}
\lref\ftwo{B. de Wit, G.L. Cardoso, D. L\"ust, T. Mohaupt and S.-J. Rey,
``Higher-order gravitational couplings and modular forms in ${\CN}=2$, $D=4$
heterotic string compactifications," Nucl. Phys. {\bf B 481} (1996) 353.}
\lref\gottsche{L. G\"ottsche, ``A conjectural generating function for numbers
of curves on surfaces," alg-geom/9711012.}
\lref\brle{J. Bryan and N.C. Leung, ``The enumerative geometry of $K3$ surfaces
and
modular forms," alg-geom/9711031.}
\lref\klm{A. Klemm, W. Lerche and P. Mayr, ``$K3$-fibrations and heterotic-type
II string duality,"
hep-th/9506122, Phys. Lett. {\bf B 357} (1995) 313.}
\lref\kutseib{D. Kutasov and N. Seiberg, ``Number of degrees of
freedom, density of states and tachyons in string theory and
CFT,'' Nucl. Phys. {\bf B 358} (1991) 600.}
\lref\khty{S. Hosono, A. Klemm, S. Theisen and S.-T. Yau, ``Mirror symmetry,
mirror map and applications to complete intersection Calabi-Yau spaces,"
hep-th/9406155, Nucl. Phys. {\bf B 433} (1995) 501.}
\lref\candelas{P. Candelas, X. de la Ossa, A. Font, S. Katz and D.R. Morrison,
``Mirror symmetry for two-parameter models-1," hep-th/9308083, Nucl. Phys. {\bf
B 416} (1994) 481; P. Candelas, A. Font, S. Katz and D.R. Morrison, ``Mirror
symmetry for two-parameter models-2," hep-th/9403187, Nucl. Phys. {\bf B 429}
(1994) 626.}
\lref\anghost{E. Witten, ``The $N$ matrix model and gauged WZW models," Nucl
Phys. {\bf B 371} (1992) 191; P. Aspinwall and D.R. Morrison, ``Topological
field theory and rational curves, " Comm. Math. Phys. {\bf 151} (1993) 245; S.
Cordes, G. Moore and S. Rangoolam, ``Lectures on 2D Yang-Mills theory,
equivariant cohomology and topological field theory," hep-th/9411210. }
\lref\hennmoore{M. Henningson and G. Moore, ``Threshold corrections in $K3
\times T^2$ heterotic compactifications, " hep-th/9608145, Nucl. Phys. {\bf B
482} (1996) 187.}
\lref\hz{J. Harer and D. Zagier, ``The Euler characteristic of the moduli space
of curves,"
Invent. Math. {\bf 85} (1986) 457.}
\lref\penner{R.C. Penner, ``Perturbative series and the moduli space of Riemann
surfaces," J. Differential Geom. {\bf 27} (1988) 35.}
\lref\mumford{D. Mumford, ``Towards an enumerative geometry of the moduli space
of curves," in M. Artin and J. Tate (eds.), {\it Arithmetic and geometry},
Birkh\"auser, 1983.}
\lref\faber{C. Faber, ``Chow rings of moduli space of curves. I. The Chow ring
of $\overline \CM_3$," Ann. Math. {\bf 132} (1990) 331.}
\lref\faberfour{C. Faber, ``Intersection-theoretical computations on $\overline
\CM_g$," in P. Pragacz (ed.), {\it Parameter spaces}, Banach Center
Publications vol. 36, Warsaw 1996.}
\lref\hoso{S. Hosono, A. Klemm, S. Theisen and S.-T. Yau, ``Mirror symmetry,
mirror map and application to Calabi-Yau hypersurfaces," hep-th/9308122, Comm.
Math. Phys. {\bf 167} (1995) 301.}
\lref\witten{E. Witten, ``Two-dimensional gravity and intersection theory on
moduli space,"
Surv. Diff. Geom. {\bf 1} (1991) 243. }
\lref\lustrev{D. L\"ust, ``String vacua with ${\cal N}=2$ supersymmetry in four
dimensions," hep-th/9803072.}
\lref\msw{J. Maldacena, A. Strominger and E. Witten, ``Black hole entropy in
M-theory," hep-th/9711053.}
\lref\lerchesti{W. Lerche and S. Stieberger, ``Prepotential, mirror map and
$F$-theory on $K3$,"
hep-th/9804176.}
\lref\hmi{Harvey, Moore, threshhold}
\lref\hmii{Harvey and Moore, Algebra of BPS states}
\lref\forgersti{K. Forger and S. Stieberger, ``String amplitudes and ${\cal
N}=2$, $d=4$ prepotential in heterotic $K3 \times T^2$ compactifications,"
hep-th/9709004, Nucl. Phys. {\bf B 514} (1998) 135.}
\lref\moore{G. Moore, ``String duality, automorphic forms, and generalized
Kac-Moody algebras," hep-th/9710198, Nucl. Phys. Proc. Suppl. {\bf 67} (1998)
56.}
\lref\kawai{T. Kawai, ``String duality and enumeration of curves by Jacobi
forms," hep-th/9804014. }
\lref\kons{M. Kontsevich, ``Intersection theory on the moduli space of curves
and
the matrix Airy function," Comm. Math. Phys. {\bf 147} (1992) 1.}
\lref\senii{A. Sen, BPS States on a Three Brane
Probe.}
\lref\si{Shioda and Inose}
\lref\mv{  Morrison and Vafa}
\lref\mfphase{E. Witten, ``Phase Transitions in
$M$-theory and $F$-theory,''  hep-th/9603150, Nucl Phys. {\bf B 471} (1996)
195.}
\lref\kleblowe{I. Klebanov and D. Lowe, ``Correlation functions in
two-dimensional quantum gravity coupled to a compact scalar field,''
Nucl. Phys. {\bf B 363} (1991) 543. }
\lref\ginsparg{P. Ginsparg and G. Moore,
``Lectures on 2D gravity and 2D string theory (TASI 1992),'' hep-th/9304011 }

\Title{\vbox{\baselineskip12pt
\hbox{YCTP-P23-98 }
\hbox{hep-th/9808131}
}}
{\vbox{\centerline{Counting higher genus curves }
\centerline{ }
\centerline{in a Calabi-Yau manifold }}
}
\centerline{Marcos Mari\~no and Gregory Moore}

\bigskip
{\vbox{\centerline{\sl Department of Physics, Yale University}
\vskip2pt
\centerline{\sl New Haven, CT 06520, USA}}
\centerline{ \it marino@waldzell.physics.yale.edu }
\centerline{ \it moore@castalia.physics.yale.edu }

\bigskip
\bigskip
\noindent
We explicitly evaluate the low energy
coupling $F_g$ in a $d=4,\CN=2$
compactification of the heterotic
string. The holomorphic piece of this
expression  provides  the information
not encoded in the holomorphic
anomaly equations, and we find that it is given
by an elementary polylogarithm with index $3-2g$,
thus generalizing in a natural way the known
results for $g=0,1$. The heterotic model has a
dual
Calabi-Yau compactification
of the type II string. We compare the
answer with the general form
expected from curve-counting
formulae and find good agreement.
As a corollary of this comparison
we predict some numbers
of higher genus curves in a specific
Calabi-Yau, and extract some
intersection numbers on the moduli space
of genus $g$ Riemann surfaces.

\Date{August 20, 1998}

\newsec{Introduction}

The computation of low energy effective actions in
supersymmetric string compactification is an interesting
problem from several points of view. First, the low energy
action summarizes many important aspects of the physics
of the compactified model. Second, the quantum corrections in
effective actions involve interesting automorphic functions.
Third, these quantum corrections often serve as generating
functions for enumerative problems in geometry. Thus, the
computation of a given physical quantity in two dual
string descriptions often leads to striking predictions
in enumerative geometry. A famous example of this is
  the   counting of rational curves in a
Calabi-Yau threefold provided by mirror symmetry. In
this paper we compute    a quantum correction involving
higher genus curves in a Calabi-Yau threefold using
heterotic/type II string duality and use the result
to make some mathematical predictions.

Specifically, we consider the well-known low energy
coupling $F_g$ of low energy effective
$d=4,\CN=2$ supergravity, introduced and  studied in
\bcov\ks\agnt. Physically, this coupling
enters the effective action in the schematic
form
\eqn\effectact{
\int F_g(t, \bar t) T^{2g-2} R^2 + \cdots
}
where $g\geq 1$, $T$ is the graviphoton field
strength, $R$ is the Riemann curvature, and
$t, \bar t$ are   vectormultiplet scalars.
If one treats $t, \bar t$ as independent
then the Wilsonian coupling is obtained by
sending $ t \rightarrow \infty$ holding
$\bar t$ fixed \ks, leaving the (anti-) holomorphic
coupling $ F_g^{\rm hol}(\bar t)$.

The mathematical formulation of $F_g$ depends on
the underlying string theory that gives rise
to the low energy supergravity. In type II compactification
on a Calabi-Yau 3-fold $X$ the expression is
exactly given by the string tree level result.
In type IIA compactification the vectormultiplet
scalars are complexified Kahler moduli and the
holomorphic coupling $\overline F_g^{\rm hol}(t)$ is
given, roughly,  by a sum over holomorphic
genus $g$ curves in $X$ \bcov\ks\agnt.
We say ``roughly'' because issues such as
curve degeneration, multiple cover formulae,
and the careful treatment of families of
curves has not yet been adequately discussed. Indeed,
one of the motivations of the present paper is
to provide some useful information for sorting
out these issues.

In heterotic compactification
on $K3 \times T^2$, the effective coupling
$F_g(t,\bar t)$ has an integral
representation coming from a one-loop diagram which
is valid to all orders of perturbation theory
\agnt. Suppose such a heterotic compactification
is dual to a type II compactification on a
Calabi-Yau $X$.
Under string duality $S_{\rm heterotic}$ is
identified with a K\"ahler class of $X$. Thus,
given a heterotic/type II dual
we can evalute a generating function for genus
$g$ curves on $X$, at one boundary of
complexified K\"ahler moduli space. This
has been done to some extent in \agnt.
In the present paper we extend the discussion
of \agnt\ by giving a complete evaluation
of $F_g$. Specifically, we
 will consider the rank four example
discussed in \kv\klm, and in many subsequent
references. We compactify the heterotic
$E_8 \times E_8$ theory on $K3 \times T^2$ and embed an $SU(2)$ bundle on
each $E_8$ with instanton number $12$. Then we Higgs completely the remaining
$E_7 \times E_7$ symmetry and we obtain a model with $244$ hypermultiplets and
the four vector multiplets corresponding to the $U(1)^4$ gauge symmetry on the
torus. In the semiclassical limit $S\rightarrow i\infty$, this is the
2-parameter $y=(T,U)$ case, with special loci corresponding to
enhanced gauge symmetries. This model is dual to a type IIA model compactified
on the
$K3$-fibered Calabi-Yau manifold $X=X_{24}^{1,1,2,8,12}$, which has
$h_{1,1}=3$, $h_{2,1}=243$, therefore $\chi (X)=-480$.

The result for
$F_g$ is naturally written as a sum of two terms
\eqn\fgeeres{
F_g = F_g^{\rm deg} + F_g^{\rm nondeg}
}
given by the rather formidable expressions equations $(4.21)$ and
$(4.40)$ below. From these expressions we can extract some
interesting results. First, the effective coupling is,
in contrast to other
quantum corrections,
continuously differentiable throughout
all of moduli space, having singularities only at the locus of
enhanced gauge symmetry $T=U$. This is hardly obvious
from the chamber-dependent evaluation of the
integral representation of $F_g$ we will give. Second,
while the expressions for $F_g$ are formidable, the
(anti-) holomorphic piece turns out to be relatively simple.
It is given by an elementary polylogarithm
${\rm Li}_{3-2g}$. The exact expression is given
 in equation $(5.2)$ below. This is our
  main result. Using the result $(5.2)$ we may draw
some conclusions about the ``number'' of genus $g$
curves in $X=X_{24}^{1,1,2,8,12}$. The variety $X$ has a
holomorphic map $\pi: X \rightarrow \IP^1$ whose
generic fibers are $K3$ surfaces. Because we must take
the limit $S\rightarrow i \infty$ we can only discuss
curves in the $K3$ fibers. Nevertheless, our result
provides some nontrivial information. The comparison
of our result for $F_g^{\rm hol}$ with known properties
of $F_g$ from \ks\ is carried out in section 6, and we
find good agreement. Moreover, we make some predictions
for numbers of genus $2$ curves in the generic
$K3$ fiber of $X$. Furthermore,    a corollary of our
discussion yields a prediction for an intersection
number on the moduli space $\CM_g$ of genus $g$
Riemann surfaces. Let $c_{g-1}$ be the Chern class
of the Hodge bundle on $\CM_g$ (see section 6 below
for a definition). Then we show that string duality
predicts:
\eqn\predicp{
\int_{\CM_g} c_{g-1}^3=(-1)^{g-1} 2
(2g-1) { \zeta (2g) \zeta (3-2g) \over
(2\pi)^{2g} }.}
This intersection number could in
principle be calculated using 2D
topological gravity, but the computation
appears to be tedious.
We expect our result $(5.2)$ to prove useful in further
investigations of the role of higher genus curves in quantum
cohomology.

Finally, we discuss briefly the method of our computation.
The integral representation of $F_g$ from the heterotic
one-loop computation was derived in \agnt\serone.
Such one-loop integrals have been evaluated in many
papers in string theory. See, for some representative examples,
\dkl\kutseib \hmalg\kawai\lerchesti. The method involves
lattice reduction and the ``unfolding technique''
(also known as the ``Rankin-Selberg method''
 in number theory, or as ``the method of
orbits.'')
The most systematic discussion of
such integrals was given by Borcherds in
\bor, generalizing the computation of
\hmalg. Because we need the more general results,
we review the notation and results of \bor\
in section three.

\newsec{The integral for $F_g$}

The $F_g$ couplings at one-loop have been computed in \serone\agnt, in the
semiclassical limit $S\rightarrow i\infty$. It is useful to introduce a
generating function for these couplings $F(\lambda,
T,U)=\sum_{g=1}^{\infty} \lambda^{2g} F_g(T,U)$. Specializing
to the heterotic dual of IIA compactification on $X_{24}^{1,1,2,8,12}$
the  integral representation of
this
generating function is
\eqn\gen{
F(\lambda, T, U)= {1 \over 2\pi^2} \int_ {\cal F} {d^2 \tau \over y} \biggl(
{E_4E_6 \over \eta^{24}} \biggr) \sum_{\Gamma^{2,2}} q^{{1\over 2}
|p_L|^2}{\overline q}^{{1\over 2}  |p_R|^2}
 \Biggl[ \biggl( { 2 \pi i  \lambda \eta^3 \over \vartheta_1 (\tilde
\lambda|\tau)} \biggr)^2 {\rm e} ^{-{\pi \tilde \lambda^2 \over y} } \Biggr].
}
In this equation, ${\cal F}$ denotes the fundamental  domain for $SL(2,
{\IZ})$, $q=\exp(2 \pi i \tau) $, $y={\rm Im}\tau$, and the modular form
$E_4E_6/\eta^{24}$ has the expansion
\eqn\cq{
{E_4 E_6 \over \eta^{24}}= \sum_{n=-1} c(n)q^n ={1\over q} -240 - \dots.}
Notice that $c(0)=\chi(X)/2$. The right and left moving momenta are
regarded as complex numbers, given by
\eqn\momenta{
\eqalign{
p_L= {1 \over {\sqrt {2 T_2 U_2}} } (n_1 + n_2 {\overline T} + m_2 U + m_1
{\overline T} U), \cr
p_R= {1 \over {\sqrt {2 T_2 U_2}}} (n_1 + n_2 T + m_2 U + m_1 TU), \cr} }
where $T=T_1 + iT_2$, $U=U_1+ i U_2$ is the decomposition into
real and imaginary parts.  Finally, $\tilde \lambda= p_R y
\lambda/{\sqrt {2 T_2 U_2}}$, and $\vartheta_1 (z|\tau)$ is the Jacobi theta
function with characteristics $(1/2, 1/2)$. We have slightly changed some of
the conventions in \serone\ in order to match the conventions of \hmalg, since
the present paper is an extension of the calculations of \hmalg. Notice
that in writing \gen\ we have momentarily introduced a term of zeroth order in
$\lambda$, as in \agnt. Our goal is to give an explicit expression for the
$F_g$ terms by performing the integral over the fundamental domain. The first
step will be to extract the $F_g$ term
from \gen\ by making an appropriate expansion of the modular forms. The
resulting integral will involve a generalized Siegel-Narain theta function with
lattice vector insertions.

The modular form involving $\vartheta_1 (z|\tau)$ can be written in terms of
Eisenstein
series:
\eqn\expon{
{ 2\pi \eta^3 z \over \vartheta_1 (z|\tau) } = -\exp \Biggl[
\sum_{k=1}^{\infty} {\zeta(2k) \over k}
E_{2k} (\tau) z^{2k}\Biggr].}
It is convenient to write \expon\ as follows. We introduce the covariant
Eisenstein series
\eqn\cov{
\hat E_2 (\tau) =E_2-{ 3 \over \pi y},
}
and the Schur polynomials:
\eqn\schur{
\exp\bigl[ \sum_{k=1}^\infty x_k z^k \bigr]
= \sum_{k=0}^\infty \CS_k(x_1, \dots, x_k) z^k,
}
which have the structure
\eqn\schurex{
\CS_k (x_1, \dots, x_k) = x_k + \dots + {x_1^k \over k!}.}
One can check that
\eqn\expansion{
\biggl( { 2 \pi i \tilde \lambda \eta^3 \over \vartheta_1 (\tilde
\lambda|\tau)} \biggr)^2 {\rm e} ^{-{\pi \tilde \lambda^2 \over y} } =
\sum_{k=0}^{\infty} \tilde \lambda ^{2k} \CP_{2k}(\hat G_2, \dots, G_{2k}),}
where we have introduced the more convenient normalized Eisenstein series
\eqn\eisen{
G_{2k} = 2 \zeta(2k) E_{2k},
}
and $\CP_{2k}$ is an almost holomorphic modular form of weight $(2k,0)$ given
by
\eqn\bigpoly{
\CP_{2k}(\hat G_2, \dots, G_{2k}) = - \CS_{k}\biggl( \hat G_2, {1\over  2} G_4,
{ 1\over
3} G_6, \dots, {1 \over  k } G_{2k}\biggr).
} We have, for instance,
\eqn\casesps{
\CP_2(\hat G_2)=-\hat G_2, \,\,\,\,\,\, \CP_4(\hat G_2, G_4)=-{1 \over 2} (\hat
G_2^2 + G_4).}

The integral \gen\ can now be written as
\eqn\reint{
F(\lambda, T,U)= \sum_{k=0}^{\infty} {\lambda^{2(k+1)} \over 2 \pi^2
(2T_2U_2)^k }
\int_{\cal F} {d^2 \tau \over y} y^{2k} {E_6 E_4 \over \eta^{24}} \CP_{2(k+1)}
\sum_{\Gamma^{2,2}}
p_R^{2k}  q^{{1\over 2} |p_L|^2}{\overline q}^{{1\over 2}  |p_R|^2}.}
In this expansion,  $k=g-1$. We see that the integrals to be computed are of
the form
\eqn\borfor{
\CI_k = \int_{\cal F} d^2 \tau  y^{2k-1} F_k(\tau) \overline \Theta,
}
where
\eqn\terms{
F_k(\tau)={E_4 E_6 \over \eta^{24}} \CP_{2(k+1)} (\hat G_2, \cdots,
G_{2(k+1)})}
is a modular form of weight $(2k,0)$ and
\eqn\thetaint{
\Theta=
\sum_{\Gamma^{2,2}} \bar p_R^{2k}  q^{{1\over 2} |p_R|^2}{\overline q}^{{1\over
2}  |p_L|^2},} is a Siegel-Narain theta function of modular weight $(1+2k,1)$.
Notice
that \borfor\ is complex except for $k=0$.

\newsec{Review of the lattice reduction and unfolding technique}

%
%
In this section we review the computation of a
  class of integrals   called theta transforms which were
systematically evaluated in \bor. The technique to
compute these integrals is to perform a lattice reduction, {\it i.e.} to reduce
the computation
to a theta transform on a smaller lattice. Proceeding iteratively, one can in
principle compute
the integral over the fundamental domain in terms of quantities associated to
the reduced
lattices. The integrals \borfor\ that we want to compute have precisely the
structure of the theta transforms considered by Borcherds, so we will briefly
review the notation and the results used in \bor\ in order to apply them to
this particular problem.

For simplicity we will restrict ourselves to self-dual lattices, although the
results in \bor\ apply to more general situations. Let $\Gamma$ be an even,
self-dual lattice
of signature $(b^+, b^-)$,  together with an isometry $P: \Gamma \otimes {\IR}
\rightarrow {\IR}^{b^+,b^-}$.
The corresponding projections on $\IR^{b^+,0}$, $\IR^{0,b^-}$ will be denoted
by $P_{\pm}$.
The inverse images of $\IR^{b^+,0}$, $\IR^{0,b^-}$ decompose $\Gamma \otimes
\IR$ into the orthogonal sum of positive definite and negative definite
subspaces. Let $p$ be a polynomial on ${\IR}^{b^+,b^-}$ of degree $m^+$ in the
first $b^+$ variables and
of degree $m^-$ in the second $b^-$ variables, and let $\Delta$ be the
(Euclidean) Laplacian in ${\IR}^{b^++b^-}$.
With these elements we construct a Siegel-Narain theta function as
\eqn\sntheta{
\Theta_{\Gamma} ( \tau; P, p)= \sum_{\lambda \in \Gamma} \exp \biggl[ -{\Delta
\over 8 \pi y }
\biggl] \bigl( p(P(\lambda)) \bigr) \exp \biggl[ \pi i \tau (P_+(\lambda))^2 +
\pi i {\overline \tau}
(P_-(\lambda))^2 \biggr] .}
One can include shifts in the lattice $\Gamma$ to obtain a more general theta
function, but
we will not  consider this case.  Notice that $(P_- (\lambda))^2 \le 0$.

As we have explained, the first step in computing the integral over the
fundamental domain, involving a theta
function with the structure \sntheta, is to perform a lattice reduction. This
is done as follows. Let $z$ be a
primitive vector of $\Gamma$ of  zero norm, choose a vector $z'$ in $\Gamma$
with
$(z,z')=1$, and let $K= (\Gamma \cap z^{\bot})/ {\IZ} z$. This lattice, which
has signature
$(b^+ + 1, b^--1)$, is the reduced lattice. The vectors in the reduced lattice
will be denoted by
$\lambda^K$, in order to distinguish them from the vectors $\lambda$ in the
original lattice. We now define ``reduced" projections $\widetilde P$ as
follows: consider $z_{\pm} \equiv P_{\pm} (z)$, and decompose ${\IR}^{b^{\pm}}
\simeq \langle z_{\pm}  \rangle \oplus \langle z_{\pm} \rangle ^{\bot}$. The
projection on the orthogonal complement $\langle z_{\pm} \rangle ^{\bot}$
is the reduced projection $\widetilde P_{\pm}$. It can be explicitly written in
terms of $P_{\pm}$ as
\eqn\redproj{
\widetilde P_{\pm} (\lambda) = P_{\pm} (\lambda) - { ( P_{\pm} (\lambda),
z_{\pm}) \over
z^2_{\pm}}  z_{\pm}.}
Once this reduced projection has been constructed, we have to decompose the
polynomial
involved in \sntheta\ with respect to this projection, according to the
expansion
\eqn\poldec{
p(P(\lambda))= \sum_{h^+, h^-} (\lambda, z_+)^{h^+} (\lambda, z_-)^{h^-}
p_{h^+, h^-} (\widetilde P (\lambda)),}
where $p_{h^+, h^-}$ are homogeneous polynomials of degrees $(m^+-h^+,
m^--h^-)$ on
$\widetilde P (\Gamma \otimes \IR )$.

Now we have to be more precise about the structure of the theta transform that
we want to compute. Consider the modular form
\eqn\modform{
F^{\Gamma} (\tau)= y^{b^+/2+m^+}F(\tau)}
with weight $(-b^-/2-m^-, -b^+/2-m^+)$, constructed from the modular form
$F(\tau)$ with weights $(b^+/2+m^+-b^-/2-m^-,0)$. We will assume that $F(\tau)$
is an almost holomorphic form, {\it i.e.} it has the expansion
\eqn\expan{
F(\tau) = \sum_{m\in \IQ} \sum_{t\ge 0} c(m,t) q^m y^{-t}
}
where $c(m,t)$ are complex numbers which are zero for all but a finite number
of values of $t$ and for sufficiently small values of $m$. In particular,
$F(\tau)$ can have a pole of finite order at cusps. The theta transform
considered in \bor\ has three ingredients: a lattice $\Gamma$ together with a
projection $P$, a polynomial  $p(P(\lambda))$, and the modular form $F(\tau)$,
and it is given by the following integral over the fundamental domain.
\eqn\thetatrans{
\Phi_{\Gamma} (P, p, F^{\Gamma}) = \int_{\cal F} {d^2 \tau \over y^2}
{\overline \Theta} (\tau;P,p) F^{\Gamma} (\tau).}
According to the results of \bor, this integral can be evaluated by reducing
the lattice. The resulting expression involves two pieces.  The first one is
essentially another theta transform but for the reduced lattice, and is given
by
\eqn\degbor{
{1 \over {\sqrt {2 z_+^2}} } \sum_{h\ge 0} \biggl( {z_ +^2 \over 4\pi }\biggr)
^h \Phi_K({\widetilde P} ,
p_{h,h} , F^K).}
The other contribution involves a sum over the reduced lattice $K$. There are
two different cases one has to consider when evaluating this contribution. When
$\widetilde P_+
(\lambda^K)$ is different form zero, one has \eqn\nondeg{
\eqalign{
&  {\sqrt {2 \over z_+^2} } \sum_{h \ge 0} \sum_{h^+, h^-} {h! \over (2i)^{h^+
+h^-}} \biggl( -{z_+^2
\over \pi } \biggr)^h {h^+ \choose h}{h^- \choose h} \sum_{j} \sum_{\lambda^K
\in K} {1 \over j!} \biggl( -{\Delta \over 8\pi}\biggr)^j \overline p_{h^+,
h^-}(\widetilde P (\lambda^K))  \cr
&\cdot  \sum_l q^{ l(\lambda^K, \mu)} \sum_t 2 c(\lambda^2/2,t) \biggl( {l
\over 2|z_+||\widetilde P_+
(\lambda^K)|} \biggr) ^{h-h^+-h^- -j -t+b^+/2+m^+-3/2}\cr
& \cdot  K_{h-h^+-h^--j-t+b^+/2+m^+-3/2}
\biggl( {2\pi l |\widetilde P_+
(\lambda^K)|\over  |z_+|} \biggr),\cr}
}
where $\mu$ is the vector in $K\otimes \IR$ given by
\eqn\muvec{
\mu= -z' + {z_+ \over z_+^2} + { z_- \over 2 z_-^2} .}
$K_\nu (z)$ is the modified Bessel function, and it comes from an integral over
the strip $y>0$.  When  $\widetilde P_+
(\lambda^K)=0$, the integral over the strip has to be regularized with a
parameter $\epsilon$. Notice that $\widetilde P_+ (\lambda^K)=0$ in two
sub-cases:
when $\lambda^K=0$, and when $\Gamma$ is a lattice with $b^+=1$,
as in this case the reduced projection will always be zero for any $\lambda^K$.
For these
situations the last sum in \nondeg\ has to be substituted by
\eqn\morered{
\eqalign{
 \sum_t  c(\lambda^2/2,t) & \biggl( { \pi l ^2\over 2z_+^2 }
\biggr) ^{h-h^+-h^-
-\epsilon-j -t+b^+/2+m^+-3/2} \cr
& \cdot \Gamma (-h+ h^++h^-+j+t-b^+/2-m^++3/2+\epsilon).\cr} }
This expression can be analytically continued to a meromorphic function of
$\epsilon$, with a Laurent expansion at $\epsilon=0$. The contribution to the
theta transform of \morered\ is given by the constant term of this expansion.
In general, the sum over $l$ will give a Riemann $\zeta$ function after
analytic continuation. In order to extract the constant term at $\epsilon =0$
one has to be careful with
possible poles in $\epsilon$ and proceed as in dimensional regularization. We
will see examples
of this in the computation of $F_g$.

An important remark concerning this result is that the theta transform will be
given by the
above expressions only for sufficiently small $z_+^2$. For a fixed primitive
vector $z$, the value of $z_+^2$ depends on the projection
we choose in our lattice, which in our case will be given by the moduli $(T,U)$
of the string
compactification. This means that the answers we will obtain for the integrals
will be chamber-dependent, {\it i.e.} they will   only be valid in a region of
moduli space. In general, to obtain the answer in some other chamber, we have
to
use wall-crossing formulae, or we must choose some other null vector
$\tilde z$ to perform the reduction in such a way that the value of
$\tilde z_+^2$ remains small in the chamber under consideration.

We will refer to \degbor\ as   the contribution of the degenerate orbit, and to
\nondeg\ as   the
contribution of the nondegenerate orbit. One should notice,
however, that the
contribution to \nondeg\ of the
zero vector of the reduced lattice appears in the computations
of \dkl\hmalg\kawai\   from the contribution of the
``degenerate orbit.''  Also, the zero orbit will appear in \degbor\
after further reduction to the trivial lattice.

\newsec{Computation of $F_g$}

We will now apply this formalism to the problem of computing the integrals
\borfor. In our case, the lattice has signature $(b^+,b^-)=(2,2)$ and is given
by
\eqn\lat{
\Gamma^{2,2} = H(-1) \oplus H(1) = \langle e_1,f_1 \rangle_{\IZ} \oplus \langle
e_2,f_2 \rangle_{\IZ} ,}
where $(e_1 , f_1) =-(e_2 , f_2)=-1$. The projections give isometries $P_{\pm}:
\Gamma^{2,2} \otimes \IR \rightarrow \IR^2$. We want to construct the
projections in such
a way that
\eqn\proj{
P_+(\lambda) = p_R, \,\,\,\,\,\,\,\ P_-(\lambda)=p_L, }
where we are identifying $\IR^2 \simeq \IC$, and $p_{R,L}$ are given as complex
numbers by \momenta. Notice that $(P_- (\lambda))^2=-|p_L|^2$. The requirement
\proj\ fixes the structure
of the projections. We can obtain explicit expressions for the projections of
the basis as follows.
Consider for instance $e_1$. As an element in the positive definite subspace of
$\Gamma^{2,2}\otimes \IR$, $P_+(e_1)$ will have the general form
\eqn\geneone{
 P_+(e_1)=x_1e_1+y_1f_1 + x_2e_2+y_2f_2 .}
Using the fact that $P_{\pm}$ are orthogonal projectors, namely $P_{\pm}^2=0$,
$P_+P_-=P_-P_+=0$, we obtain the equations
\eqn\projeq{
x_1+x_2 \overline T+ y_2 U + y_1 \overline T U =0, \,\,\,\ x_1+x_2  T+ y_2 U +
y_1 T U =1.}
Solving for $x_1$, $x_2$, $y_1$, $y_2$ we obtain
\eqn\projex{
P_+(e_1) = {1 \over 2T_2U_2} \biggl\{
-{\rm Re} (TU) e_1 -f_1+U_1 e_2 +T_1 f_2 \biggr\} ,}
and similarly
\eqn\projexf{
P_+(f_1) =  {1 \over 2T_2U_2} \biggl\{
-|TU|^2 e_1 -{\rm Re} (TU) f_1+|U|^2T_1 e_2 +|T|^2U_1 f_2 \biggr\}.
}
Consider now the theta function in \terms. The Laplacian in terms of the
variables $p_R$, $p_L$ is given by
\eqn\lapl{
\Delta= 4 \bigl( {d \over dp_R} {d \over d\overline p_R} +{d \over dp_L} {d
\over d\overline p_L}
\bigr),}
therefore $\Delta (\overline p_R^{2k} ) =0$ and we see that the theta function
involved in the
one-loop integral is just the Siegel-Narain theta function for the projection
given by \proj\projex\
and polynomial $p=(\overline p_R)^{2k}$, which is homogeneous of degree $2k$ in
the $\IR^{b^+}$ variables. Therefore, $m^+=2k$, $m^-=0$. For each $k$ we have
an integral
\borfor\ with the structure \thetatrans\ and
\eqn\fmods{
F_k^{\Gamma}(\tau)=y^{2k+1} F_k(\tau),}
where $F_k(\tau)$ is given in \terms. We will denote the coefficients in the
expansion
\expan\ by $c_k(m,t)$. Notice that, in this expansion, the terms of the form
$y^{-t}$ come from the $\hat G_2$ in the Schur polynomials. Therefore,
according to \schurex, the range of $t$ is $0 \le t \le k+1$.

To perform the lattice reduction, we have to choose a primitive null vector in
$\Gamma^{2,2}$.
A natural choice is $z=e_1$, $z'=-f_1$. It is easy to check that
\eqn\checknorm{
(z_+, \lambda) = {\sqrt {z_+^2}} {\rm Re}(p_R),}
where
\eqn\znorm{
z_+^2= { 1 \over 2T_2U_2}.}
According to our remarks in section 3, the answer obtained with the lattice
reduction will be valid
for $1\ll T_2U_2$.  This choice of $z$ is convenient for the decomposition
in \poldec, as one has
\eqn\decomcon{
(\overline p_R )^{2k} =({\rm Re} (p_R)-i {\rm Im} (p_R))^{2k} =
\sum_{h^+=0}^{2k} (z_+, \lambda)^{h^+} p_{h^+,0}(\lambda),
}
where the reduced polynomials are given by
\eqn\decomagain{
 p_{h^+,0}(\lambda)= {2 k \choose h^+} {(-i)^{2k-h^+} \over |z_+|^{h^+}} ({\rm
Im}(p_R))^{2k-h^+}.}
The reduced lattice is $K=\langle e_2,f_2 \rangle$, and the reduced projection
can be easily obtained from \redproj\ and \checknorm:
\eqn\redpex{
\widetilde P_+ (\lambda)= {\rm Im}(p_R), \,\,\,\,\,\ \widetilde P_- (\lambda)=
{\rm Im}(p_L).}
We will now compute the integrals using the expressions \degbor\nondeg.

\subsec{The degenerate orbit}

First we compute the contribution of the degenerate orbit, which is now a theta
transform for the lattice $K$, with vectors of the form
\eqn\redvectors{
p_R^K={1 \over {\sqrt { 2T_2 U_2}}} (n_2 T +m_2 U), \,\,\,\,\,\  p_L^K={1 \over
{\sqrt { 2T_2 U_2}}} (n_2 {\overline T} +m_2 U).}
Notice that the lattice $K$ is Lorentzian, with $(b_+, b_-)=(1,1)$. According
to \degbor, the new theta transform involves the lattice $K$ (together with the
projection $\widetilde P$) and the polynomial $p_{0,0}(\widetilde P(\lambda^K))
= (-1)^k ({\rm Im}(p^K_R))^{2k}$ (remember from \decomagain\ that $h^-=0$),
which is again homogeneous of degrees $(m^+,m^-)=(2k,0)$. The modular forms
involved in this theta transform are then, according to \modform,
\eqn\modredlat{
F_k^K(\tau) =y^{2k +1/2}F_k(\tau).}

To evaluate the theta transform for $K$, we have to perform a further
reduction to the trivial lattice, using for instance the vector $\hat z=e_2$,
with
\eqn\newsq{
\hat z_+^2= { T_2 \over 2 U_2}.}
The norm of this vector is computed using the reduced projection ${\widetilde
P}$. The polynomial appearing in the new theta function can be decomposed again
(with respect to the new projection associated to $\hat z$) using \poldec:
\eqn\newredpol{
(-1)^k ({\rm Im}(p^K_R))^{2k}={(-1)^k \over |\hat z_+|^{2k}} (\lambda^K, \hat
z_+)^{2k},}
therefore the reduced polynomial is a constant $\hat p_{2k,0}=(-1)^k/|\hat
z_+|^{2k}$. For the new theta transform we apply the result of section 3 again,
and we have two contributions, corresponding to degenerate and nondegenerate
orbits for the reduction of $K$ to the trivial lattice. The degenerate orbit is
simply a theta transform \degbor\ for the trivial lattice. Because of the
structure of the reduced polynomial, only the $k=0$ integral will give a
nonzero contribution. From the definition  \thetatrans\ we then have that this
theta transform is given by
\eqn\zerored{
\Phi_0(\cdot, 1, F_0)= \int_{\CF} {d^2 \tau \over y^2}{E_4 E_6 \over \eta^{24}}
\CP_2 (\hat G_2) = 16 \pi^3,}
where we have used that $\CP_2 (\hat G_2)=-\hat G_2$ and the general result
\latst\
\eqn\lsw{
\int_{\CF} {d^2 \tau  \over \tau_2^2} \bigl( \hat G_2 (\tau) \bigr)^n F(\tau) =
{1 \over \pi (n+1) } \biggl[ \bigl( G_2 (\tau))^{n+1} F(\tau) \biggr]_{q^0}.}
The theta transform for the zero lattice corresponds in fact to the
contribution of
the $A=0$ orbit in the
usual unfolding technique.

To evaluate the contribution of the nondegenerate orbit, we have to use the
expression \morered, as the reduced lattice is now trivial. The sum $\sum_{l>0}
l^{2(1+t-k)-\epsilon}$ can be analytically continued to the Riemann zeta
function
$\zeta (2(1+\epsilon+t-k))$, and it is easy to check that there are no poles at
$\epsilon=0$ (recall that the only pole of $\zeta(z)$ is at $z=1$).
Taking all this into account one obtains,
\eqn\degorbit{
\CI_k^{\rm deg} = 16 \pi   U_2 \delta_{k,0}+ {1 \over 2^{k-1}}
\sum_{t=0}^{k+1} c_k(0,t) {t!\over \pi^{t+1} } T_2 \biggl( {T_2 \over
U_2} \biggr)^{t-k} \zeta (2(1+t-k)).}
Including the rest of the factors, and recalling that $k=g-1$, we obtain the
contribution of the degenerate orbit to $F_g$:
\eqn\fgdeg{
F_g^{\rm deg}= 8 \pi^3 U_2 \delta_{g,1} +{ 1 \over
T_2^{2g-3}} \sum_{t=0}^{g} c_{g-1} (0,t) {t!\over 2^{2(g-1)} \cdot \pi^{t+3} }
\biggl( {T_2 \over
U_2} \biggr)^t \zeta (2(2+t-g)).}
Due to our choice of $\hat z$ and to \newsq, the expression \degorbit\ is only
valid in the chamber $T_2 < U_2$. The result in the other chamber is obtained
by interchanging $T_2 \leftrightarrow U_2$ in the above formula.

\subsec{The nondegenerate orbit}
Let's now compute the contribution of the non-degenerate orbit. We will denote
the vectors in the reduced lattice $K$ by $r \equiv \lambda^K=ne_2 + mf_2$. We
first define the following products:
\eqn\products{
r\cdot y = nT + mU, \,\,\,\,\,\  r\hat{\cdot} y = {\rm Re} (nT + mU) + i \vert
{\rm Im} (nT + mU)|.}
The action of
the Laplacian on the reduced polynomial $\bar p_{h^+,0}(r)$ in \decomagain\
gives the sum
\eqn\laplaction{
\sum_{j=0}^{[k-h^+/2]} \sum_{r\in K} {(-1)^j i^{2k-h^+}  \over (8\pi )^j
|z_+|^{h^+ }}  {(2k)! ({\rm Im}(p^K_R))^{2k-h^+-2j}  \over j! (2k-h^+ -2j)!
h^+! }. }
We have to consider the two different cases $r=0$, $r\not=0$. In the first
case, we have to use the expression \morered. Notice that \laplaction\ vanishes
when $r=0$, unless $2k-h^+=2j$. This forces $h^+$ to be even, say $h^+=2s$,
where $0\le s \le k$, and
$j=k-s$. We then find, for the contribution of the zero vector,
\eqn\zerovec{
\eqalign{
\CI_{k, r=0}^{\rm nondeg} =& 2 \sum_{s=0}^{k} \sum_{t=0}^{k+1} (-1)^s
2^{2(s-2k)+t} { (2k)! \over (2s)! (k-s)!} {c_k(0,t) \over \pi^{t+1/2}}
|z_+|^{2(t-k)} \cr
& \cdot \biggl( {\pi \over 2z_+^2}\biggr)^{-\epsilon} \Gamma ({1\over 2}+
\epsilon + s+t -k)
\zeta (1+ 2\epsilon +2(t-k)),\cr}}
where the sum over $l$ has been analytically continued again to a Riemann
$\zeta$-function. According to Borcherds's formula, we have to evaluate the
constant term of the Laurent expansion of this expression around $\epsilon=0$.
The only possible pole is in the Riemann
$\zeta$-function, and occurs when $t=k$. Otherwise the above expression is
analytic at
$\epsilon =0$. To evaluate the constant term at $\epsilon=0$ when $t=k$, we
expand in $\epsilon$ as in dimensional regularization, using:
\eqn\epsilonexp{
\eqalign{
\zeta (1 + 2\epsilon)=&{1 \over 2\epsilon} + \gamma_{\rm E} +
{\CO}(\epsilon),\cr
 \Gamma({1\over 2}+ s+ \epsilon) = &\Gamma({1\over 2}+s) \bigl[ \psi({1\over 2}
+s) + \epsilon \bigr] + {\CO}(\epsilon^2),\cr
 \biggl( {\pi \over 2z_+^2}\biggr)^{-\epsilon}=&1-\epsilon \log \biggl( {\pi
\over 2z_+^2}\biggr)+
  {\CO}(\epsilon^2),\cr}}
where $\gamma_{\rm E}$ is the Euler-Mascheroni constant and $\psi (z)$ is the
logarithmic derivative of the $\Gamma$ function. The constant term in the
Laurent expansion can be easily evaluated using the above expressions. To write
the final answer, one notices that some of the sums on $s$ that appear in the
result can be explicitly computed, using $\Gamma (1/2+n)= \pi^{1/2}2^{-n}
(2n-1)!!$ and $\Gamma (1/2-n)= (-1)^n \pi^{1/2}2^{n} / (2n-1)!!$, with $n\ge
0$. We have for instance,
\eqn\sumons{
 \sum_{s=0}^{k}  (-1)^s 2^{2s} { (2k)! \over (2s)! (k-s)!} \Gamma
(1/2+s)=\pi^{1/2} \delta_{k,0},}
and the results turn out to be very different for $k=0$ and $k \not=0$. For
$k=0$ we have,
\eqn\finalzero{
 \CI_{k=0, r=0}^{\rm nondeg} =c_0 (0,0) \biggl[ -\log  \biggl( {\pi \over
2z_+^2}\biggr) + \gamma_{\rm E} -2 \log 2 \biggr] + c_0(0,1) {2 \zeta (3) \over
\pi} |z_+|^2,}
and for $k\not=0$,
\eqn\finalnon{
\eqalign{
 \CI_{k \not=0, r=0}^{\rm nondeg} = & 2  \sum_{t=0}^{k-1}  {c_k(0,t) \over
\pi^{t+1/2}} \zeta (1+2(t-k))
|z_+|^{2(t-k)} \cr
& \cdot \sum_{s=0}^{k} (-1)^s 2^{2(s-2k)+t} { (2k)! \over (2s)!
(k-s)!}\Gamma ({1\over 2}+  s+t -k) \cr
&+ {c_k(0,k) \over 2^{3k} \cdot \pi^k}
\sum_{s=0}^{k}  (-1)^s { (2k)! \over s! (k-s)!} \psi( {1\over 2} + s).\cr
}}

For nonzero vectors in $K$, we use \nondeg. Taking into account that $(\lambda,
\mu)={\rm Re}( r\hat{\cdot} y)$, we obtain:
\eqn\nondegex{
\eqalign{
\CI^{\rm nondeg}_{k, r\not=0}&=  {\sqrt {2 \over z_+^2} }\sum_{h^+=0}^{2k}
\sum_{j=0}^{[k-h^+/2]} \sum_{r\not= 0 } {(-1)^{j+k+h^+}  \over (8\pi )^j (2
|z_+|)^{h^+ }}  {(2k)! ({\rm Im}(p_R))^{2k-h^+-2j}  \over j! (2k-h^+ -2j)! h^+!
} \cr
&\cdot  \sum_{l=1}^{\infty}   \sum_{t=0}^{k+1}  2 c_k(r^2/2,t) \biggl( {T_2U_2
l \over {\rm Im} (r\hat{\cdot} y)} \biggr) ^{\nu} l^{h^+} {\rm e} ^{2 \pi i l
{\rm Re}( r\hat{\cdot} y) }K_{\nu}
\bigl( 2\pi l {\rm Im}(r\hat{\cdot} y) \bigr),\cr}
}
In this equation, $\nu=2k-h^+-j-t-1/2$. Notice that this is always a
half-integer, therefore we can use the explicit expression for the modified
Bessel function when $s\ge 0$
\eqn\bessel{
K_{s+1/2} (x)= {\sqrt { \pi \over 2 x}} {\rm e}^{-x} \sum_{k=0}^{s} {(s+k)!
\over k! (s-k)!} {1 \over (2x)^n}.}
We will give now our final expression for the contribution of the nondegenerate
orbit. We define $s$ as $s+1/2=|\nu|$, and we take into account that
$K_s(x)=K_{-s}(x)$. We also introduce the polylogarithm
function,
\eqn\polylog{
{\rm Li}_m(x) = \sum_{\ell=1}^\infty  {x^\ell \over  \ell^m}.
}
Notice that for $m\leq 0 $ these functions are elementary: ${\rm
Li}_0(x)=x/(1-x)$, and
\eqn\polylogi{
{\rm Li}_{m}(x) = \bigl(x {d \over  dx} \bigr)^{\vert m \vert } {1 \over  1-x}
= m! { x^{\vert m \vert} \over  (1-x)^{\vert m \vert +1} }
+ \cdots, \,\,\,\,\ m<0 } We then see that \nondegex\ can be written as:
\eqn\ansii{
\eqalign{
\CI^{\rm nondeg}_{k, r\not=0}&= 2^{1-k} \sum_{r\not= 0}
 \sum_{t=0}^{k+1}
 \sum_{h=0}^{2k} \sum_{j=0}^{[k-h/2]}
\sum_{a=0}^s c_k(r^2/2, t)
{(-1)^{j+h+k} \over  (4 \pi)^{j+a} }
{(2k)! \over  j! h! (2k-h-2j)!}
{(s+a)! \over  a! (s-a)!} \cr &  \cdot ({\rm sgn} ({\rm Im} (r\cdot y))^h
(T_2U_2)^{k-t} ({\rm Im} [r \hat{\cdot} y])^{t-j-a}
{\rm Li} _{1+a+j+t-2k}({\rm e} ^{2 \pi i r \hat{\cdot} y} ).
\cr}
}

We will now begin to simplify the complicated expression
\ansii. The nature of the sum depends strongly
 on the sign of ${\rm
Im}(r\cdot y)$. If ${\rm Im}(r\cdot y) >0$ then \ansii\ {\it vanishes}
for $g\geq 3$. The reason is that there is an
unconstrained sum on $h$. For $g=1$ we get:
\eqn\zerok{
\CI^{\rm nondeg}_{0, r\not=0}=2 \sum_{r\not= 0} \biggl[ c_0(r^2/2,0){\rm Li}_1
({\rm e}^
{2\pi i r \hat{\cdot} y})+ {c_0(r^2/2,1)\over
T_2 U_2 } \CG (r \hat {\cdot} y) \biggr],
}
where the function $\CG (x)$ is defined as:
\eqn\mypoly{
\CG (x) = {\rm Im}(x) {\rm Li}_2 ({\rm e}^
{2\pi i x}) + {  1\over 2\pi }{\rm Li} _3 ({\rm e}^
{2\pi i x}).}
For $g=2$ and ${\rm Im}(r\cdot y) >0$ we get:
\eqn\onek{
\CI ^{\rm nondeg}_{1, r\not=0}=-{   c_1(r^2/2,2)\over
  \pi^2 T_2U_2}{\rm Li}_3 ({\rm e}^
{2\pi i r \hat{\cdot} y}).
}
When ${\rm Im}(r\cdot y) <0$ we get more complicated sums.
For $g=2$ we have a sum involving polylogarithms ${\rm Li}_k$ with index
$-1\le k \le 3$, but for $g\geq 3$  the sum starts with a rational function.
The structure
of \ansii\ is, for ${\rm Im}(r\cdot y) <0$ and $g>2$:
\eqn\struc{
\sum_{t=0}^{g}c_{g-1} (r^2/2,t) (T_2 U_2)^{g-1-t} f_t ({\rm Im}(r \hat{\cdot}
y), {\rm Li}_n ({\rm e}^{2\pi i r \hat{\cdot} y})),
}
where the polynomials $f_t$ are of the following form
\eqn\polyn{
f_t= \sum_{p=0}^{\Lambda}a_{p}^{(t)} ({\rm Im} (r \hat{\cdot} y))^{t-p} {\rm
Li}_{t+p+3-2g}({\rm e}^
{2\pi i r \hat{\cdot} y}).}
In this equation, $a_p^{(t)}$ are some numerical constants and $\Lambda= {\rm
min}\{ t, 2g-t-3 \}$. This implies that the index of the polylogarithms is
always less than or equal to zero.

Putting all the contributions together, we find that the contribution to $F_g$
from the nondegenerate orbit is then, for $F_1$
\eqn\fonenondeg{
\eqalign{
F_1^{\rm nondeg} =& {1 \over \pi^2} \sum_{r\not= 0} \biggl[ c_0(r^2/2,0){\rm
Li}_1 ({\rm e}^
{2\pi i r \hat{\cdot} y})+ {c_0(r^2/2,1)\over
T_2 U_2 } \CG (r \hat {\cdot} y) \biggr]\cr
& +{c_0 (0,0)\over 2\pi^2}  \biggl[ -\log (T_2U_2)  + \gamma_{\rm E} -2 \log
2\pi \biggr] + {c_0(0,1) \zeta (3) \over 2\pi^2 (T_2U_2)}.\cr}}
It is interesting to notice that $F_1$ is essentially given by the integral
$\widetilde \CI_{2,2}$ of \hmalg. If we explicitly compute the contribution of
the degenerate orbit \fgdeg\ for $g=1$ (which gives a piecewise polynomial in
$T_2$, $U_2$), we find that $F_1$ agrees with the result presented in \hmalg,
after taking into account the different normalizations.
Finally, for $g>1$ we find
\eqn\fgnondeg{
\eqalign{
&F_{g>1}^{\rm nondeg} = {(-1)^{g-1} \over 2^{2(g-1)} \cdot \pi^2} \sum_{r\not=
0 }
 \sum_{t=0}^{g}
 \sum_{h=0}^{2g-2} \sum_{j=0}^{[g-1-h/2]}
\sum_{a=0}^s c_{g-1} ( r^2/2, t)
{(2g-2)! \over  j! h! (2g-h-2j-2)!}
 \cr &  \cdot  {(-1)^{j+h} \over  (4 \pi)^{j+a} }{(s+a)! \over  a! (s-a)!}({\rm
sgn} ({\rm Im} (r\cdot y))^h
{1\over (T_2U_2)^t} ({\rm Im} [r \hat{\cdot} y])^{t-j-a}
{\rm Li} _{3+a+j+t-2g}({\rm e} ^{2 \pi i r \hat{\cdot} y} )
\cr
& +{c_{g-1}(0,g-1) \over 2^{4g-5} \cdot \pi^{g+1}} {1\over (T_2 U_2)^{g-1}}
\sum_{s=0}^{g-1}  (-1)^s { (2g-2)!  \over s! (g-1-s)!} \psi ({1\over 2}+s)  \cr
& + \sum_{t=0}^{g-2} {c_{g-1}(0,t) \over \pi^{t+5/2}} {\zeta (3+2(t-g)) \over
(T_2U_2)^t}  \sum_{s=0}^{g-1}(-1)^s 2^{2(s-2g+2)} { (2g-2)! \over (2s)!
(g-1-s)!}   \Gamma ({3\over 2}+  s+t -g) .\cr}}
According to \terms\ and \expan, the coefficients $c_{g-1} (m,t) $ are given by
the expansion
\eqn\expex{
{E_4 E_6 \over \eta^{24}} \CP_{2g} (\hat G_2, \cdots, G_{2g})= \sum_{m\in \IQ}
\sum_{t\ge 0} c_{g-1} (m,t) q^m y^{-t},}
where the polynomials $\CP_{2g}(\hat G_2, \cdots, G_{2g})$ are defined in
\bigpoly.
The coefficients $c_{g-1}(-1,0)$ and $c_{g-1}(0,0)$ will be important
in the next sections, so we will determine them. As $t=0$ for these
coefficients, they are found by looking at the holomorphic part of \expansion.
Using the
representation of $\vartheta_1(z|\tau)$ as an infinite product, it is easy to
check that
\eqn\qexpquo{
-\biggl( { 2\pi \eta^3 z \over \vartheta_1 (z|\tau) }\biggr)^2=- \biggl( {\pi z
\over \sin \pi z} \biggr)^2 + 8\pi^2 z^2 q + {\cal O} (q^2).}
This implies that
\eqn\qone{
\bigl[ \CP_2 (G_2) \bigr]_q=8\pi^2,  \,\,\, \bigl[ \CP_{2g} (G_2, \dots,
G_{2g}) \bigr]_q=0, \,\, g>1. }
On the other hand, as $E_4E_6/\eta^{24} =q^{-1} -240+ \dots$, the coefficient
$c_{g-1}(-1,0)$ is determined by the $q^0$ term of $\CP_{2g}(G_2, \cdots,
G_{2g})$, and therefore can be computed using the expansion
\eqn\coneexp{
\biggl( {\pi z  \over \sin \pi z} \biggr)^2= -\sum_{g=0}^{\infty} {(2g-1)!
\over (2g)!} (-1)^g (2 \pi z)^{2g} B_{2g},}
where $B_{2g}$ are the Bernoulli numbers.
We then obtain, using the relation between $\zeta(2g)$ and $B_{2g}$,
\eqn\cg{
c_{g-1}(-1,0)=-2(2g-1) \zeta (2g).}
In the same way, using \qone, we find
\eqn\chiq{
c_{g-1}(0,0)=c_{g-1}(-1,0) {\chi (X) \over 2}, \,\,\, g>1,}
where $X=X_{24}^{1,1,2,8,12}$ is the Calabi-Yau manifold on the type IIA side.

As the pole at the cusp in \expex\ is of first order, $c_{g-1}(m,t)$ will be
zero for $m<-1$. This implies that, in the expression \fgnondeg, the most
negative possible value of $r^2/2$ is $-1$. Therefore, the integers $n$, $m$ in
\products\ giving
a nonzero contribution are $n\ge 0$, $m \ge 0$, $n \le 0$, $m\le 0$ and the two
lattice points
$(n,m)=(1,-1)$, $(n,m)=(-1,1)$. As $T_2$, $U_2>0$, the hatted dot product that
appears in \fgnondeg\ will give a chamber structure only for the two points in
the lattice with $nm=-1$, and the walls will be defined by $T_2=U_2$.

In general, when discussing theta transforms, one has to make
an analytic continuations of the polylogarithms through
the wall $T_2=U_2$, and there is a nonzero wall-crossing term. An important and
remarkable property of $F_g$ that is not obvious from \fgdeg\ and \fgnondeg\ is
that the
wall-crossing of the degenerate orbit will exactly cancel the wall-crossing of
the nondegenerate
orbit. In other words, {\it the whole coupling $F_g$ is continuously
differentiable on the moduli
space} with coordinates
$(T,U)$, for $g>1$, except at the locus of enhanced gauge symmetry
$T=U$ where
there is a singularity.  This
cancellation of wall-crossing seems to require a precise knowledge of many
terms in the sums
\fgdeg\ and \fgnondeg, as well as the precise values of the coefficients
$c_{g-1}(0,t)$ and $c_{g-1} (-1,t)$, and we have checked it numerically up to
genus $5$. Note that this is physically reasonable since there are
no massless singularities at the generic point on this wall. This should
be contrasted with the situation in five dimensions.
Upon decompactification to five dimensions the wall $T_2=U_2$
 corresponds to
the location of massless particle singularities. Considered as
a   $K3 \times S^1$ compactification these are well-known enhanced
symmetry particles at the self-dual radius of $S^1$.  Considered
as an M-theory compactification on a Calabi-Yau manifold
the masless particles are due to wrapped M2-branes \aft\mfphase.

\newsec{Extracting the holomorphic piece}

Although the expressions just obtained are somewhat intimidating,
 it turns out
that
the antiholomorphic part of $F_g$ has a very simple and compact expression.
Notice that, with our choices for
$p_{L,R}$ in \momenta, the piece of $F_g$ which does not mix the holomorphic
and antiholomorphic parts is in fact antiholomorphic. Hence, for simplicity
of presentation,  we will
state our results for ${\overline F}_g$, and the holomorphic piece of this
function of the moduli will be denoted by $\overline F_g^{\rm hol}$. This piece
should have
a geometrical interpretation in terms of counting of holomorphic curves on the
target space,
according to \bcov\ks. To extract $\overline F_g^{\rm hol}$, with $g\ge 2$, we
take the holomorphic limit $\overline T$, $\overline U \rightarrow \infty$ in
${\overline F}_g$.

The first thing to notice is that
the degenerate orbit does not give any contribution in this limit, for $g\ge
2$.
We therefore
 consider the behavior of the nondegenerate orbit, for $g\geq 2$. In the
first two lines of \fgnondeg, one can easily see that the only surviving term
in the holomorphic limit has $j=t=a=0$, and the resulting function is then
${\rm Li}_{3-2g}(x)$. In order to write a closed expression for this term, we
have to be more precise about the contributions of the reduced lattice $K$.
 As ${\rm Im} (r\cdot y) >0$ for the lattice
points $n\ge 0$, $m \ge 0$, there is no holomorphic contribution from these for
$g\ge 2$. The lattice point $(1,-1)$ only contributes when $T_2<U_2$, giving
${\rm Li}_{3-2g} (q)$, where $q=\exp (2\pi i (T_1-U_1)
+2\pi (T_2-U_2))$. The point $(-1,1)$ only contributes for $T_2>U_2$, giving
${\rm Li}_{3-2g}(q^{-1})$. Using
\polylogi\ for $m < 0$, one can easily check that
\eqn\noflop{
{\rm Li}_m (x) = (-1)^{|m|+1} {\rm Li}_m (x^{-1}).}
Therefore, the contributions
of $(1,-1)$ and of $(-1,1)$ add up to ${\rm Li}_{3-2g}(q)$ for any value of
$T_2$ and $U_2$. Finally, the terms in the last two lines of \fgnondeg, which
correspond to the
contribution of $r=0$, vanish as $\overline T$, $\overline U \rightarrow
\infty$, except for the term with $t=0$ in the last line. The sum over $s$ in
the last factor can be easily evaluated to be $(-1)^{g-1}\pi^{1/2}/2$. We then
obtain, taking into account \cg\ and \chiq,
\eqn\holo{
\overline F^{\rm hol}_g= {(-1)^{g-1} \over \pi^2} \biggl[ -(2g-1)\zeta (2g)
\zeta (3-2g){\chi (X)\over 2}   +  \sum_{r>0} c_{g-1}(r^2/2,0) {\rm Li}_{3-2g}
({\rm e}^
{2\pi i r {\cdot}y})\biggr],
}
for $g\ge 2$. In this equation, $r>0$ means the following possibilities:
$(n,m)=(1,-1)$, $n > 0$, $m>0$,  $n=0$, $m>0$, or $n>0$, $m=0$. Notice that the
coefficients $c_{g-1}(r^2/2,0)$ involved in $F^{\rm hol}_g$ are determined by
the holomorphic modular form $\CP_{2g} (G_2, \dots, G_{2g})$.

Let us once again consider the nature of the answer at the wall
$T_2=U_2$. We note from \noflop\ that
there is no wall-crossing behavior at
$T_2=U_2$ for $\overline F^{\rm hol}_g$, $g\ge 2$. This is of course expected,
as we are taking a particular limit of the whole $F_g$ which does not have any
wall-crossing (as we have checked for $g\leq 5$). For
$\overline F^{\rm hol}_g$, $g\ge 2$, one can indeed prove it analytically using
\noflop.  Viewed from the type IIA perspective the
absence of wall-crossing is consistent with the
fact that the Hodge numbers, and therefore the Euler character, of
$X$ are preserved under flop transitions.

In contrast to the absence of wall-crossing there is a singularity
in codimension two at $T=U$.
One can easily deduce the structure of the leading singularity in $\overline
F^{\rm hol}_g$ using our explicit answer. The
singularities on the heterotic side correspond to the appearance of extra
massless states
on the $T=U$ locus, and on the type II side they correspond to the conifold
singularity. The enhanced gauge symmetry occurs when $p_L^2=2$, $p_R^2=0$,
{\it i.e.} when $(n,m)=(1,-1)$, so that $r^2/2=-1$. The leading singularity at
genus $g$ in the expression \holo\ is given by the leading pole in the
polylogarithm, which has the form
\eqn\sing{
 {(-1)^{g-1} \over \pi ^2} c_{g-1}(-1,0)  (2g-3)! {1 \over (1-{\rm e}^{2\pi i r
{\cdot} y})^{2g-2}}.}
Notice that, for $r {\cdot}y \rightarrow 0$,  we find the expected singular
behavior $(2\pi i r {\cdot} y)^{-(2g-2)}$. The coefficient of this leading term
can be computed explicitly: taking into account the factor $(2g-3)!$ in \sing,
and the expression for $c_{g-1}(-1,0)$ in \cg, we see that it is given by
\eqn\leadcoef{
-2^{2g} \pi^{2g-2}\chi (\CM_g),}
where $\chi (\CM_g)=B_{2g}/2g(2g-2)$ is the Euler characteristic of the moduli
space of genus $g$ Riemann surfaces \hz\penner.

Up to a normalization factor, we recover
the expansion of the free energy for the $c=1$ string at the self-dual radius,
as suggested by the conjecture of
 \gv\ and verified by \agnt\   for a similar model. Notice that, according to
our result, this pole together with the subleading poles of $F_g$ near the
conifold locus add up to the function
${\rm Li}_{3-2g}$. It is also worth noting in this context that one
can combine the result \holo\ together with \expansion\polylogi\
to produce a compact expression for the sum $\sum_g \lambda^{2g} \bar F_g^{\rm
hol}$,
essentially given by
\eqn\sumup{
\sum_{r>0} \Biggl[ q^{-r^2/2} {E_4 E_6 \over \eta^{24}}
\biggl( { 2 \pi i \tilde \lambda \eta^3 \over \vartheta_1 (\tilde
\lambda|\tau)} \biggr)^2\Biggr]_{q^0} {\rm Li}_3(x)\Biggr\vert_{x=e^{2 \pi i
r\cdot y}}
}
where we now take $\tilde \lambda = \lambda x{d\over dx}$. This expression
is highly reminiscent of the formula in $c=1$ theory giving the
radius dependence of the free energy \kleblowe\ginsparg.

\newsec{Counting higher genus curves}

Using string duality and the results in \bcov\ks, the $\overline F_g^{\rm hol}$
terms that we have computed should count numbers of curves of genus $g$ on the
$K3$-fibered Calabi-Yau $X$, on the type IIA side. This has been checked to
some extent in \fone\ for $\overline F_1^{\rm hol}$. We will analyze here the
$\overline F_2^{\rm hol}$ term
in some detail, and make some observations on the higher $\overline F_g^{\rm
hol}$.

Unfortunately, not   much is known about the structure of the $F^{\rm
IIA}_g$ in   type IIA  theory for $g>1$.
We do know   the contribution to $F^{\rm IIA}_g$ coming from constant maps,
which was obtained in \ks\ for any $g>1$. This is the term in $F^{\rm IIA}_g$
that survives in the limit in which the K\"ahler moduli $t$, $\bar t$ go to
infinity, and it involves an integral over the moduli space of Riemann surfaces
$\CM_g$. Let $\CH$ be the Hodge bundle on $\CM_g$, whose fibre over the point
$m \in \CM_g$ is given by $H^0 (\Sigma_g (m), K)$, where $K$ is the canonical
bundle of the Riemann surface
$\Sigma_g (m)$. Let $c_k=c_k (\CH)$ be their Chern classes. The
constant term of
$F^{\rm IIA}_g$ is then given by
\eqn\fglimit{
F^{\rm IIA}_g \big|_{t, \bar t \rightarrow \infty} = {1 \over 2} \chi (X)
\int_{\CM_g} c_{g-1}^3,}
where $\chi (X)$ is the Euler characteristic of the Calabi-Yau
threefold $X$. The value of the above integral over the moduli space of Riemann
surfaces is known to be $1/2880$ for $g=2$ \mumford, $1/725760$ for $g=3$
\faber, and $1/43545600$ for $g=4$ \faberfour. As we will
see, string duality gives a very precise prediction for this coefficient for
any genus $g>1$.

In the case of $F^{\rm IIA}_2$, although we don't have a general expression,
explicit results have been obtained in \ks\ for some Calabi-Yau manifolds.
These examples suggest that $F^{\rm IIA}_2$ has the following structure:
\eqn\ftwoks{
F^{{\rm IIA}} _2 ( {\rm e}^{2\pi i l {\cdot}t}) = {\chi (X)\over 5760}
-\chi(\CM_2) \sum_l  d_l
 {    {\rm e}^{2\pi i l {\cdot}t}
\over (1- {\rm e}^{2\pi i l {\cdot} t})^2} + \sum_l D_l  {\rm e}^{2\pi i l
{\cdot} t},}
In this equation, $d_l$ counts the number of rational curves, and $D_l$ counts
the
number of holomorphic curves of genus $2$. The
$t=(t_1, \dots, t_{h_{1,1}})$ denotes complexified K\"ahler moduli, $l=(l_1,
\dots, l_{h_{1,1}})$ are integers, and
\eqn\newprodk{
l{\cdot} t =\sum_{i=1}^{h_{1,1}} l_it_i.}
The first term in \ftwoks\ is the contribution to $F^{\rm IIA}_2$ from constant
maps \fglimit, and $-\chi(\CM_2)=1/240$ is dictated by the structure of the
leading singularity
at the conifold. In order to compare the result on the heterotic side with the
type IIA expression,
we have to take into account that we are working in the semiclassical limit
$S\rightarrow i\infty$. This corresponds on the type IIA side to the region of
the K\"ahler cone where the volume of the
base of the $K3$-fibration goes to infinity, {\it i.e} to the limit in which
the K\"ahler modulus
dual to $S$ goes to infinity. Therefore, our result \holo\ will take into
account only
holomorphic curves in the fiber of the Calabi-Yau.

 The first thing to notice is that the function involved in the first member of
\ftwoks\ is
precisely ${\rm Li}_{-1} ( {\rm e}^{2\pi i l {\cdot} y})$. We will now match
our main result \holo\ to \ftwoks, and extract some predictions for the numbers
$D_l$. This can be done in a very precise way taking into account that,
according to the results of \hmalg, the number of rational curves on the
K3 fiber on the
type IIA side is counted by $-2c(r^2/2)$, where the coefficients $c(n)$ are
defined by \cq. In order to extract the genus $2$ counting function, we write
$\CP_4 (G_2, G_4)$ as
\eqn\split{
-{1 \over 2} (G_2^2 + G_4)= -{\pi^4 \over 15} + {\pi^4 \over 15} \biggl(
{6-5E_2^2 -E_4 \over 6} \biggr),}
where we used that $\zeta (2)=\pi^2/6$, $\zeta (4)= \pi^4/90$. It is convenient
to define the coefficients $\widetilde c (n)$ as follows,
\eqn\coefftwo{
{1 \over 720} {E_4E_6  \over  \eta^{24} }(6-5E_2^2-E_4 )   =
\sum_{n=1}^{\infty}
\widetilde c (n) q^n =6q - 1408 q^2 - 856254 q^3- \dots.}
It follows from the definition of $c_{g-1}(n, 0)$ that
\eqn\splitc{
c_{g-1}(n, 0) = -{\pi^4 \over 15} \biggl( c(n) -120 \widetilde c(n) \biggr) ,}
for $g=2$.
The holomorphic coupling $\overline F^{\rm hol}_2$ can then be written,
according to \holo,  as \foot{
An explicit expression for the holomorphic piece of $F_2$ was obtained in
\ftwo\ using the anomaly equation together with
target space duality. Although we have not performed a detailed comparison of
the two expressions, the result presented in \ftwo\ seems to involve also the
polylogarithm function
${\rm Li}_{-1} ({\rm e}^{2\pi i r {\cdot} y})$. }
\eqn\ftwobis{
\overline F^{\rm hol} _2=-8 \pi^2 \biggl[ -{480\over 5760} + {1 \over 240}
\sum_{r>0} (-2c(r^2/2))  { {\rm e}^{2\pi i r {\cdot} y}
\over (1- {\rm e}^{2\pi i r {\cdot} y})^2} +
\sum_{r>0} \sum_{k=1}^{\infty} \widetilde c(r^2/2) k {\rm e}^{2\pi i k r
{\cdot} y}
\biggr] . }
The first term of this expression comes from the constant term in \holo, and
remarkably agrees with the type IIA side result in \ftwoks. The second term
corresponds to the contribution of rational curves in $F_2$, and has the
structure found in \ks. The remaining piece corresponds to  genus two curves on
the $K3$-fibered Calabi-Yau. We can now extract some predictions for the genus
$2$ instanton numbers. The Calabi-Yau $X$ has three K\"ahler moduli, denoted by
$t_1$, $t_2$ and $t_3$ (with the notation of \hoso\fone), and the heterotic
weak coupling corresponds to $t_2 \rightarrow \infty$. This means that we will
only be able to compute the instanton numbers $D_{l_1,0,l_3}$. The relation
between the remaining K\"ahler moduli and the
heterotic moduli is $t_1=U$, $t_3=T-U$ \klm\fone, therefore $l_1=n+m$, $l_3=n$.
Using this explicit map, and assuming that \ftwoks\ is in fact the right
structure on the type IIA side, we find for instance (for the very first values
of $l_1$, $l_3$)
\eqn\numbers{
\eqalign{
& D_{l,0,l}=D_{0,0,l}=D_{1,0,l}=0,\cr
&D_{2,0,1}=6, \,\,\,\ D_{3,0,1}=D_{3,0,2} =-1408,\cr
&D_{4,01}=D_{4,0,3}=-856254, \,\,\,\  D_{4,0,2}=-55723284, \cr
&D_{5,0,1}=D_{5,0,4}=-55723296,\cr
& D_{5,0,2}=D_{5,0,3}=-34256077056,\cr }}
and so on. For primitive $\ell_1, \ell_3$ and
$\ell_1-\ell_3 \gg 1, \ell_3\gg 1$ we have the asymptotic
result
\eqn\asympt{
D_{\ell_1,0,\ell_3} \sim -{1 \over 120 \sqrt{2}} ((\ell_1 -\ell_3)\ell_3)^{1/4}
\exp\bigl[
4 \pi \sqrt{(\ell_1 -\ell_3)\ell_3 }\bigr].
}

Notice that all the instanton numbers will be integer numbers, due
to the form of the expansion \coefftwo, but most of them are negative. The same
phenomenon was observed in many two-parameter models \candelas\khty, as well as
in this particular Calabi-Yau \fone, for the genus one instanton numbers. The
reason for this is that the curves with genus $g>1$ come in families, and what
we are really computing is the Euler character of a certain vector bundle over
the moduli space of curves in the family. In the context of topological sigma
models, this vector bundle is associated to the antighost zero modes \anghost.

We can try to follow the same strategy at arbitrary genus. In particular, we
should be able to
recover the limiting behavior \fglimit\ of $F_g^{\rm IIA}$ by fixing the
normalizations. This can be done if we consider a natural generalization of
\ftwobis\ and write $F^{\rm hol}_g$ as:
\eqn\fgbis{
\eqalign{
\overline F_g^{\rm hol} = -2(2\pi)^{2g-2} \biggl[ &(-1)^{g-1}{2 (2g-1) \zeta
(2g) \zeta (3-2g) \over (2\pi)^{2g}} {\chi (X) \over 2}\cr
& -\chi(\CM_g) \sum_{r>0} (-2 c(r^2/2,0)){({\rm e}^{2\pi i r {\cdot} y})^{2g-3}
\over (1-{\rm e}^{2\pi i r {\cdot} y})^{2g-2}} + \cdots\biggr],\cr}}
where we have decomposed $c_{g-1}(r^2/2,0)=c_{g-1}(-1,0) c(r^2/2)+ \cdots$, and
written only the leading term of the polylogarithm in \polylogi. We then see
that the relative normalization of $F_g$ is given by the factor
$-2(2\pi)^{2g-2}$. Matching the constant term appearing here to \fglimit, we
obtain
\eqn\predic{
\int_{\CM_g} c_{g-1}^3=(-1)^{g-1} 2 (2g-1) { \zeta (2g) \zeta (3-2g) \over
(2\pi)^{2g} },}
and one is startled to find
the correct values $1/725760$ for $g=3$ and $1/43545600$ for $g=4$. Notice that
for
$g=0$ we recover
precisely the well-known constant term  of the prepotential $-\zeta (3)\chi
(X)/2$.

\newsec{Conclusions}
We have found an elegant expression for the holomorphic part of $F_g$ in terms
of
a polylogarithm function, as well as an explicit expression for the generating
function of the
number of curves of genus $g$ in terms of a modular form of weight $2g-2$.
Unfortunately, the structure of $F_g$ on the type IIA side, for $g>2$,  has not
been explored in detail, and this makes more difficult a geometrical
interpretation
of our expressions, as well as a precise prediction of the genus $g$ instanton
numbers. Notice that the information about the genus $g$ curves is contained
in the modular form $F_{g-1} (\tau)$ defined in \terms, but in order to extract
it one needs a previous knowledge of the different contributions to $F_g$, as
we have seen
in genus $2$ . It is interesting to notice that our counting function has some
similarity with the modular form that counts genus $g$ curves with $g$ nodes on
a $K3$ surface \gottsche\brle.
On the other hand, the fact that we have an exact and simple expression for
$F^{\rm hol}_g$
should be very helpful in trying to understand the geometrical interpretation
of higher genus
curve counting.

We have also made two concrete predictions: first, assuming the structure of
$F_2$
given by \ftwoks\ (as suggested by the examples considered in \ks) we have
obtained the modular form that gives the instanton numbers at genus $2$ on the
Calabi-Yau manifold
 $X_{24}^{1,1,2,8,12}$. Some of them have been written in \numbers. But we also
have
a prediction with a different flavor in \predic, where a certain intersection
number on the moduli space of genus $g$ curves has been obtained from string
duality. Both should be testable with other methods already available. The
instanton numbers could be obtained on the
type IIA side by mirror symmetry, following the original strategy of \ks. The
general form of the result obtained here will certainly help in fixing the
holomorphic ambiguity. As for the relation
\predic, one can perhaps make further checks using the approach to
two-dimensional
topological gravity presented in \witten\kons\ or by standard methods in
algebraic geometry as in \mumford\faber\faberfour.

Although we have only performed the computation for a single model,
similar calculations could be done in other situations, such
as those considered
in \hennmoore. Since the integral representation for $F_g$ will
be quite similar to that considered here, we expect that
the simple form for $F_g^{\rm hol}$ in terms of a polylogarithm function
also holds
in more general situations.

\bigskip
\centerline{\bf Acknowledgements}\nobreak
\bigskip

We would like to thank A. Gerasimov, S. Katz, A. Losev, D. Morrison,
R. Plesser,  and S. Shatashvili for
useful discussions, and M. Serone for useful correspondence. GM would like
to thank the Aspen Center for Physics for hospitality during the completion
of this paper. This work  is supported by
DOE grant DE-FG02-92ER40704.

\listrefs

\bye